\documentclass[useAMS,usenatbib]{mn2e}
\usepackage{graphicx}
\DeclareGraphicsExtensions{.eps,.cps,.ps,.eps.gz,.ps.gz,.eps.Z}
\title[{GRB~050802.}]{The Two-Component Afterglow of {\it Swift} GRB~050802}

\author[]{S. R. Oates$^{1}$, M. De Pasquale$^{1}$, M.J. Page$^{1}$, A.J. Blustin$^{1}$, S. Zane$^{1}$, K. McGowan$^{2}$,\newauthor K.O. Mason$^{1}$, T.S. Poole$^{1}$, P. Schady$^{1}$, P.W.A. Roming$^{3}$,K.L. Page$^{4}$, A. Falcone$^{3}$,\newauthor N. Gehrels$^{5}$\\
$^{1}$ Mullard Space Science Laboratory, University College London,
Holmbury St. Mary, Dorking Surrey, RH5 6NT, UK \\
$^{2}$School of Physics and Astronomy, University of Southampton, Highfield, Southampton, SO17 1BJ\\
$^{3}$Department of Astronomy and Astrophysics, Pennsylvania State University, 525 Davey Laboratory, University Park, PA 16802. \\
$^{4}$Department of Physics and Astronomy, University of Leicester, University Road, Leicester LE1 7RH, UK \\
$^{5}$NASA Goddard Space Flight Center, Laboratory for High Energy Astrophysics, Greenbelt, MD 20771\\}

\date{Released 2002 Xxxxx XX}

\begin{document}

\date{Accepted...Received...}

\maketitle

\label{firstpage}

\begin{abstract}
This paper investigates GRB~050802, one of the best examples of a {\it Swift} gamma-ray burst afterglow that shows a break in the X-ray lightcurve, while the optical counterpart decays as a single power-law. This burst has an optically bright afterglow of 16.5 magnitude, detected throughout the 170\,-\,650\,nm spectral range of the UVOT on-board {\it Swift}. Observations began with the XRT and UVOT telescopes $\rm 286\,s$ after the initial trigger and continued for $\rm 1.2\,\times\,10^6\,s$. The X-ray lightcurve consists of three power-law segments: a rise until 420\,s, followed by a slow decay with $\alpha_2$\,=\,0.63\,$\pm$\,0.03 until 5000\,s, after which, the lightcurve decays faster with a slope of $\alpha_3$\,=\,1.59\,$\pm$\,0.03. The optical lightcurve decays as a single power-law with $\alpha_{O}$\,=\,0.82\,$\pm$\,0.03 throughout the observation. The X-ray data on their own are consistent with the break at 5000\,s being due to the end of energy injection. Modelling the optical to X-ray spectral energy distribution, we find that the optical afterglow can not be produced by the same component as the X-ray emission at late times, ruling out a single component afterglow. We therefore considered two-component jet models and find that the X-ray and optical emission is best reproduced by a model in which both components are energy injected for the duration of the observed afterglow and the X-ray break at 5000\,s is due to a jet break in the narrow component. This bright, well-observed burst is likely a guide for interpreting the surprising finding of {\it Swift} that bursts seldom display achromatic jet breaks.

\end{abstract}


\section{INTRODUCTION}
\label{intro}

Gamma Ray Bursts (GRBs) are the most energetic explosions that take place in our Universe, with a typical energy of $\rm10^{51}\,-\,10^{53}\,ergs$ released on a timescale of between a millisecond and a few thousand seconds. The release of such a considerable amount of energy over such a short period requires an outflow that is relativistic \citep{mes97} and is likely to be anisotropic \citep{sar99}. The energy within the outflow is released primarily though shocks. The GRB is thought to be produced through internal shocks from interactions between successive shells of ejecta \citep{sar97}. Subsequently an afterglow is emitted as the outflow is decelerated through collisionless shocks with the external medium \citep{sar97a}. A forward shock propagates into the external medium and emits from X-ray to radio wavelengths, and a reverse shock travels back through the ejecta and peaks at longer wavelengths than the forward shock \citep{zhaf05}. 

The afterglow can reveal many properties of the progenitor and its surroundings. Currently, the most up to date model \citep{zhaf05} allows the use of the temporal and spectral indices to indicate the nature of the surrounding medium, whether it is a uniform density medium, or a medium with a density that is decreasing radially as expected for a stellar wind e.g $ n \propto r^{-2}$, where $n$ is the particle density and $r$ is the radius from the GRB. The indices also indicate the location of the observing band relative to the synchrotron self absorption frequency $\nu_a$, the peak frequency $\nu_m$ and cooling frequency $\nu_c$.

Some GRB afterglows have a period of slow decline in their lightcurves ($F(t)\propto t^{-\alpha}$, where $\alpha\sim 0.5$) \citep{zhaf05}, which is generally accepted to be due to continued energy injection \citep{rees98}. This energy could be due to a central engine that is long lasting \citep{dai98, zha01}, later shells catching up and colliding with slower shells that were emitted earlier \citep{rees98}, or the slow release of energy stored in the form of Poynting Flux \citep{zha05}. The end of energy injection is signalled by an increase in decay rates of the afterglow lightcurves to $F(t)\propto t^{-\alpha}$, where $\alpha\sim1$.

Currently the most effective observatory for studying GRBs and their early afterglows is {\it Swift}, which has now been in operation for over two years. It has the ability to observe emission ranging from $\gamma$-rays to optical with the three on-board telescopes, namely the Burst Alert Telescope (BAT; \citeauthor{bar05} 2005), the X-ray Telescope (XRT; \citeauthor{bur05} 2005) and the Ultra Violet and Optical Telescope (UVOT; \citeauthor{rom05} 2005). The BAT covers a large area of the sky allowing of order 100 bursts to be detected per year. When the BAT has been triggered by the GRB, the satellite slews automatically allowing the XRT and the UVOT to begin observing as soon as possible, usually within $\rm 100\,s$ of the burst trigger. {\it Swift} is thus able to observe both the initial $\gamma$-ray explosion and the following early afterglow.

This paper looks at GRB~050802, a burst that appears to challenge the standard picture. At the end of the shallow decay the X-ray afterglow breaks to a steeper decay, while the optical afterglow continues to decay as a power-law without a break. If both the X-ray and optical arise from the same component and if the X-ray break were due to the end of energy injection then the optical lightcurve would be expected to break at the same time. Such behaviour has been noted recently in 6 {\it Swift} GRB afterglows \citep{pan06} of which GRB~050802 is currently the best example. This burst has a X-ray lightcurve with two distinct breaks and it was observed for 1.2\,$\times\,10^6$\,s. It also had an optically bright afterglow ($\sim$16 magnitude at early times) that was well sampled in 6 filters of the UVOT up to 1~$\times~10^{5}$~s and thereafter observed with the white UVOT filter. In this paper we discuss the possible models that could explain this behaviour. We will use the convention flux $ F\,=\,t^{-\alpha}\,\nu^{-\beta}$ with $\alpha$ and $\beta$ being the temporal and spectral indices respectively. We assume the Hubble parameter $ H_0\,=\,70$\,$\rm km\,s^{-1}\,Mpc^{-1}$ and density parameters $\Omega_\Lambda$\,=\,0.7 and $\Omega_m$\,=\,0.3. Uncertainties are quoted at 1$\sigma$ unless otherwise stated.
 
\section{{\it Swift} and Ground Based Observations}

The BAT was triggered by GRB~050802 at 10:08:02~UT on the $2^{nd}$~August~2005 \citep{ban05}. The lightcurve rises for 5\,s to the first of three peaks and has a $T_{90}\,=\,(13\,\pm\,2)$\,s (90 \% confidence level). The fluence in the $\rm 15\,-\,350\,keV$ band is $\rm (2.8\,\pm\,0.1)\,\times\,10^{-6}\,erg\,cm^{-2}$ to $90\,\%$ confidence level \citep{pal05}. Observations with the XRT and UVOT began 289\,s and 286\,s respectively, after the BAT trigger\citep{ban05,mcg05}. Both the XRT and UVOT continued to observe until $\rm 1.2\,\times\,10^{6}\,s$ after the burst trigger.

The XRT began observations by locating the burst with Image Mode (IM). After the burst was located, data were taken in Windowed Timing (WT) mode for 163\,s. A fading uncatalogued source was found within $8^{''}$ of the BAT position \citep{ban05} and was confirmed as the X-ray counterpart of GRB~050802. 480\,s after the burst trigger, the XRT changed modes and continued observations in Photon Counting (PC) mode. 

UVOT observations showed a fading, uncatalogued source at RA\,=\,$14^{h}$\,$37^{m}$\,$05.69^{s}$, Dec\,=\,$27^{\circ}$\,$47^{'}$\,$12.2^{''}$ \citep{mcg05}. Following the trigger, a series of automated exposures were taken in the three optical and three ultra-violet filters. A log of the observations is given in Table \ref{tab:uvotpoints}. The observations consisted of an initial 100\,s `finding chart' exposure in the V-band, 10\,s exposures in each passband for 7 rotations of the filter wheel, followed by a sequence of $\sim 100$\,s and $\sim 900$\,s exposures. Later observations (after $1\,\times\,10^5$\,s), were taken in the UVOT white filter. 
      
The afterglow was also imaged and detected with the 2.6\,m Shain Telescope, 8\,hrs after the burst, with R and i' band magnitudes of 20.6 and 20.2 respectively \citep{pav05}. Spectroscopic observations were carried out with ALFOSC on the Nordic Optical Telescope. Several absorption features were detected providing a redshift measurement of $z\,=\,1.71$ (\citeauthor{fyn05} 2005a $\&$ 2005b). \cite{fyn05} discovered an extended source within 1'' of the UVOT afterglow location, which they propose could be the host galaxy. 

\section{XRT and UVOT data reduction.}
\label{XUdata}
\subsection{X-ray Data}

The XRT data were reduced using the XRT pipeline software V2.0. Source and background counts were taken from the cleaned event files using extraction regions, in order to construct spectra and lightcurves. Events of grade 0\,-\,2 were used for the WT mode and 0\,-\,12 were used for the PC mode. For the WT mode data we used a 40 pixel strip for the source extraction region and a 40 pixel strip for the background extraction region. In the PC mode data the first 2.5\,ks of data were found to be piled-up and so required the use of an annular extraction region. The size of the region that was affected by pile up was determined by comparing the radial profile of the afterglow with a model of the XRT point spread function (PSF). The radial profile and the model PSF are inconsistent at radii less than 2.5 pixels ($6\arcsec$), so we used an inner radius of 2.5 pixels and an outer radius of 30 pixels ($71\arcsec$). The size of the source extraction region for the rest of the PC data was 30 pixel ($71\arcsec$) radius. For all PC data, background counts were extracted from a circular region of radius 80 pixels ($189\arcsec$). Appropriate response matrices (RMs) were taken from the Swift calibration database, CALDB 20060424 and effective area files  were constructed using the standard XRT software. A correction factor was calculated and applied to the piled-up section of the XRT lightcurve to account for the excluded, piled up pixels. Fortunately, the source was not located near the bad pixel columns and so no correction was required.

\subsection{UVOT Data}

The UVOT event files were screened for bad times (e.g. South Atlantic Anomaly passage, Earth-limb avoidance) and the images were corrected for Mod-8 noise using the standard UVOT software. The images were transformed to sky coordinates and then corrected for the $\sim\,5\arcsec$ uncertainty in the aspect of the spacecraft pointing using bespoke software.

Counts for the afterglow were extracted using an aperture of radius $4\arcsec$ for the optical filters and $5\arcsec$ for the ultra-violet filters. Background subtraction was performed using counts extracted from a larger region offset from the source position. The measured count rates were aperture-corrected to radii of $6\arcsec$ for the V, B and U filters and 12$\arcsec$ for the UV filters. These were then translated to magnitudes using the standard UVOT zero-points (see Table \ref{tab:uvotpoints}).

The optical lightcurve is shown in Fig. \ref{fig:lateopticallc}. The lightcurve from each filter of the UVOT was normalized to the V filter. The normalization factor for each filter (see Table \ref{tab:uvotnorm}) was calculated by taking the average count rate through the filter in the 400s\,-\,1000s time range, and dividing this value by the average count rate in the V filter over the same time interval. Later observations were obtained with the UVOT white filter; these were normalized to the equivalent V count rates as follows. The optical-UV spectral energy distribution (SED) was modelled using the average count rates from the V, B, U, UVW1, UVM2 and UVW2 filters in the 400s\,-\,1000s time range. The optical/UV response matrices were then used to predict the ratio of the V to white count rates.

\subsection{Combined X-ray and UV/optical Spectral Energy Distributions}
SEDs were produced spanning the optical to X-ray range for early (400s\,-\,1000s) and late (35ks\,-\,55ks) times. For each period, the average count rates of the exposures in each UVOT filter were used to produce the optical spectral values. For the X-ray part of each SED a spectrum was extracted in the relevant time range.

\begin{table*}
{\scriptsize
\begin{tabular}{|l c c c  c l l c c c c } 
\hline
Filter & $\rm T_{Mid}$ (s) & Exposure & Count Rate & Magnitude &  & Filter & $\rm T_{Mid}$ (s) & Exposure & Count Rate & Magnitude\\
\hline                                                                               
V &      291  &    10    &3.5  	         $\pm$0.8	  &$16.5^{+0.3}_{-0.2} $ & \vline  & UVW1 &      578  &  10	&0.55   $\pm$0.36                &  $18.5^{+1.2}_{-0.6}$   \\	        	
V &      301  &    10    &1.8  	         $\pm$0.7	  &$17.2^{+0.5}_{-0.4} $ & \vline  & UVW1 &      662  &  10	&0.76   $\pm$0.39                &  $18.1^{+0.8}_{-0.4}$	 \\   		      
V &      311  &    10    &1.7   	 $\pm$0.7	  &$17.2^{+0.6}_{-0.4} $ & \vline  & UVW1 &      747  &  10	&0.97   $\pm$0.42                &  $17.9^{+0.6}_{-0.4}$	 \\    		      
V &      321  &    10    &3.3   	 $\pm$0.8	  &$16.5^{+0.3}_{-0.2} $ & \vline  & UVW1 &      831  &  10     &0.26   $\pm$0.26                &  $19.3^{+4.5}_{-0.7}$	 \\    		      
V &      331  &    10    &0.98           $\pm$0.60	  &$17.9^{+1.0}_{-0.5} $ & \vline  & UVW1 &      915  &  10	&0.40   $\pm$0.29                &  $18.8^{+1.4}_{-0.6}$	 \\    		      
V &      341  &    10    &2.7  	         $\pm$0.8	  &$16.8^{+0.4}_{-0.3} $ & \vline  & UVW1 &      1045  &  100   &0.44   $\pm$0.09                &  $18.7^{+0.3}_{-0.2}$	 \\    		      
V &      351  &    10    &1.9  	         $\pm$0.7	  &$17.1^{+0.5}_{-0.3} $ & \vline  & UVW1 &      1671  &  100   &0.52   $\pm$0.10                &  $18.5^{+0.2}_{-0.2}$	 \\    		      
V &      361  &    10    &3.0  	         $\pm$0.8	  &$16.6^{+0.3}_{-0.3} $ & \vline  & UVW1 &      2296  &  100   &0.21   $\pm$0.07                &  $19.5^{+0.5}_{-0.3}$	 \\    		      
V &      371  &    10    &2.1  	         $\pm$0.7	  &$17.0^{+0.4}_{-0.3} $ & \vline  & UVW1 &      2905  &  66    &0.10   $\pm$0.08                &  $20.3^{+1.7}_{-0.6}$	 \\    		      
V &      381  &    10    &2.6  	         $\pm$0.7	  &$16.8^{+0.4}_{-0.3} $ & \vline  & UVW1 &      14220  & 578    &0.079   $\pm$0.026             &  $20.6^{+0.4}_{-0.3}$	 \\    		      
V &      466  &    10    &2.2  	         $\pm$0.7	  &$17.0^{+0.4}_{-0.3} $ & \vline  & UVW1 &      31454  &  823  &0.041   $\pm$0.020              &  $21.3^{+0.7}_{-0.4}$	 \\    		      
V &      550  &    10    &1.1            $\pm$0.6	  &$17.8^{+0.9}_{-0.5} $ & \vline  & UVW1 &      43037  &  796  &0.005   $\pm$0.019              &$<$20.8 (3$\sigma)$ \\      	      	
V &      634  &    10    &0.93  	 $\pm$0.57	  &$17.9^{+1.0}_{-0.5} $ & \vline  & UVW1 &      54604  &  803  &0.012   $\pm$0.020              &$<$20.7 (3$\sigma)$	 \\   		
V &      718  &    10    &2.4  	         $\pm$0.7	  &$16.9^{+0.4}_{-0.3} $ & \vline  & UVM2 &     395     & 10    &-0.054  $\pm$0.130              & $<$18.4 (3$\sigma)$ \\	       	
V &      803  &   10     &0.91           $\pm$0.49	  &$17.9^{+0.8}_{-0.5} $ & \vline  & UVM2 &     480     & 10    &0.19   $\pm$0.18                &  $19.0^{+3.2}_{-0.7}$	 \\    	
V &      888  &   10     &2.0  	         $\pm$0.6	  &$17.1^{+0.4}_{-0.3} $ & \vline  & UVM2 &     564     & 10    &0.073   $\pm$0.130              &  $<$18.0 (3$\sigma)$	 \\    	       	
V &      972  &   10     &0.76          $\pm$0.44	  &$18.1^{+0.9}_{-0.5} $ & \vline  & UVM2 &     648     & 10    &0.16   $\pm$0.18                &  $<$17.6 (3$\sigma)$	 \\   	       	
V &      1464  &  10     &0.90  	 $\pm$0.14	  &$17.9^{+0.2}_{-0.2} $ & \vline  & UVM2 &     732     & 10    &0.33   $\pm$0.22                &  $18.4^{+1.1}_{-0.5}$  	 \\    	
V &      2088  &  10     &0.53         $\pm$0.12	  &$18.5^{+0.3}_{-0.2} $ & \vline  & UVM2 &     817     & 10    &-0.055  $\pm$0.130              & $<$18.4 (3$\sigma)$	 \\    	       	
V &      2714  &  100    &0.33         $\pm$0.11 	  &$19.0^{+0.5}_{-0.3} $ & \vline  & UVM2 &     901     & 10    &0.056   $\pm$0.130              &$<$18.1 (3$\sigma)$ \\    	       	
V &      12566  & 100    &0.11        $\pm$0.04 	  &$20.2^{+0.5}_{-0.3} $ & \vline  & UVM2 &     986     & 10    &-0.055  $\pm$0.130              & $<$18.4 (3$\sigma)$ \\               
V &      25528  & 100    &0.028        $\pm$0.032   	  &$<$20.1 (3$\sigma$)  &  \vline  & UVM2 &     1567      &  100   &0.073   $\pm$0.041   &   $20.0^{+0.9}_{-0.5}$	 \\            	
V &      41283  & 900   &-0.044      $\pm$0.041	          &$<$20.6 (3$\sigma$) &   \vline  & UVM2 &     2192      &  100   &0.0051   $\pm$0.0284       &$<$19.8 (3$\sigma)$ 	 \\    	       	
V &      52838  & 900    &0.055        $\pm$0.041	  &$21.0^{+1.4}_{-0.6} $ & \vline  & UVM2 &     2818      &  100   &0.038   $\pm$0.038      & $<$19.2 (3$\sigma)$	 \\   	       	
B &      437  &    10  &4.6  	 $\pm$1.0       	  &$17.5^{+0.3}_{-0.2} $ & \vline  & UVM2 &     13473     &  900   &0.021   $\pm$0.010      & $21.4^{+0.7}_{-0.4}$    \\     	       	
B &      521  &    10  &3.3  	 $\pm$0.9	          &$17.8^{+0.3}_{-0.3} $ & \vline  & UVM2 &     26032   &  95   &-0.019  $\pm$0.024      & $<$20.4 (3$\sigma)$	 \\   		       	
B &      605  &    10  &2.8  	 $\pm$0.9                 &$18.0^{+0.4}_{-0.3} $ & \vline  & UVM2 &     30585     &900  &0.004   $\pm$0.009       &$<$20.9 (3$\sigma)$	 \\    	     	       	
B &      690  &    10  &3.4  	 $\pm$0.9	          &$17.8^{+0.3}_{-0.2} $ & \vline  & UVM2 &     42188     &898  &-0.005  $\pm$0.009       &$<$21.4 (3$\sigma)$ \\    	      	       	
B &      774  &    10  &4.1  	 $\pm$0.9	          &$17.6^{+0.3}_{-0.2} $ & \vline  & UVM2 &     53461   &  861  &0.004   $\pm$0.010       &$<$20.8 (3$\sigma)$	 \\    	      	       	
B &      859  &    10  &4.3  	 $\pm$0.9	          &$17.5^{+0.2}_{-0.2} $ & \vline  & UVW2 &     452       &10  	&-0.151  $\pm$0.133     &  $<$19.3 (3$\sigma)$	 \\	      	       	
B &      943  &    10  &3.7  	 $\pm$0.8	          &$17.7^{+0.3}_{-0.2} $ & \vline  & UVW2 &     537     &  10  	&0.329   $\pm$0.273     & $19.0 ^{+1.9}_{-0.7}$ \\	       	       	
B &      1254  &   100 &1.9  	 $\pm$0.2	          &$18.4^{+0.1}_{-0.1} $ & \vline  & UVW2 &     621     &  10  	&0.023   $\pm$0.137      & $<$18.7 (3$\sigma)$	 \\   		       	
B &      1879  &   100 &1.5  	 $\pm$0.2	          &$18.7^{+0.1}_{-0.1} $ & \vline  & UVW2 &     705       &10  	&-0.036  $\pm$0.134      & $<$18.9 (3$\sigma)$	 \\     	       	
B &      2505  &   100 &0.80   $\pm$0.16                  &$19.4^{+0.2}_{-0.2} $ & \vline  & UVW2 &     790     &  10   &0.26   $\pm$0.22     &  $19.3^{+2.2}_{-0.7}$	 \\    	     	       	
B &      7631  &   900 &0.38  $\pm$0.05	                  &$20.2^{+0.1}_{-0.1} $ & \vline  & UVW2 &     874     &  10 	&0.40   $\pm$0.26     &  $18.8^{+1.1}_{-0.5}$	 \\      	       	
B &      36405  &  900&0.20   $\pm$0.05	                  &$20.9^{+0.3}_{-0.2} $ & \vline  & UVW2 &     958       &  10      &0.028   $\pm$0.13     &  $<$18.8 (3$\sigma)$	 \\   	       	
B &      47974  &  900&0.085    $\pm$0.043	          &$21.8^{+0.8}_{-0.4} $ & \vline  & UVW2 &     1359      &  100	&0.036   $\pm$0.038     &  $<$19.9 (3$\sigma)$	 \\            	
U &      423  &  10    &4.2  	 $\pm$0.8	          &$16.8^{+0.2}_{-0.2} $ & \vline  & UVW2 &     1984      &  100	&0.039   $\pm$0.043     &  $<$19.8 (3$\sigma)$   \\            	
U &      507  &  10    &4.7  	 $\pm$0.9	          &$16.7^{+0.2}_{-0.2} $ & \vline  & UVW2 &     2610      &  100	&0.063   $\pm$0.043     & $20.8^{+1.2}_{-0.6}$	\\  	       	
U &      591  &  10    &2.3  	 $\pm$0.7	          &$17.5^{+0.4}_{-0.3} $ & \vline  & UVW2 &     8406      &  634  &0.028   $\pm$0.015     & $21.7^{+0.8}_{-0.5}$	\\     	       	
U &      675  &  10    &3.3  	 $\pm$0.8	          &$17.0^{+0.3}_{-0.2} $ & \vline  & UVW2 &     20050     &  272  &-0.034  $\pm$0.018      & $<$22.0 (3$\sigma)$	\\     	       	
U &      760  &  10    &2.6  	 $\pm$0.7	          &$17.3^{+0.3}_{-0.2} $ & \vline  & UVW2 &     24621     &  900  &-0.016  $\pm$0.012      & $<$22.1 (3$\sigma)$	\\     	       	
U &      844  &  10    &3.3  	 $\pm$0.7	          &$17.1^{+0.3}_{-0.2} $ & \vline  & UVW2 &     37257     &  786  &-0.013  $\pm$0.012      & $<$21.9 (3$\sigma)$	\\     	       	
U &      929  &  10     &2.4  $\pm$0.6	                  &$17.4^{+0.3}_{-0.2} $ & \vline  & UVW2 &     48827   &  789  &-0.028  $\pm$0.011     &$<$23.6 (3$\sigma)$	\\     	      	       	
U &      1149  & 100 &1.9  	 $\pm$0.2	          &$17.6^{+0.1}_{-0.1} $ & \vline  & WHITE &    441766  & 10950 &  0.02  $\pm$0.04    &  $<$21.9 (3$\sigma)$    \\      	      	       	
U &      1775  & 100 &1.4  	 $\pm$0.2	          &$18.0^{+0.1}_{-0.1} $ & \vline  & WHITE &    611955 &  3417 &   -0.14 $\pm$0.08    &  $<$22.3 (3$\sigma)$  	\\     	      	       	
U &      2401  & 100 &1.0      $\pm$0.1	                  &$18.3^{+0.2}_{-0.1} $ & \vline  & WHITE &    696516 &  3595 &   -0.11 $\pm$0.06    &  $<$22.6 (3$\sigma)$ \\     	      	       	
U &      6723  & 900  &0.47  	 $\pm$0.04	          &$19.2^{+0.1}_{-0.1} $ & \vline  & WHITE &    783075 &  5179 &   -0.06 $\pm$0.05    &  $<$22.3 (3$\sigma)$	\\     	      	       	
U &      35498  &900  &0.17    $\pm$0.04	          &$20.3^{+0.3}_{-0.2} $ & \vline  & WHITE &    869866 &  4507&    -0.08 $\pm$0.06    &  $<$22.1 (3$\sigma)$ 	\\     	      	       	
U &      47066  &900  &0.11    $\pm$0.04	          &$20.8^{+0.5}_{-0.3} $ & \vline  & WHITE &    995723 &  3645 &   -0.16 $\pm$0.09    &  $<$22.2 (3$\sigma)$     \\     	      	       	
UVW1 &   409  &  10    &1.5    $\pm$0.5	                  &$17.4^{+0.5}_{-0.3} $ & \vline  & WHITE &    1.21e+06 &1712 &   -0.10  $\pm$0.14    &  $<$21.0 (3$\sigma)$ \\	       	      	       	
UVW1 &   494  &  10   &1.6     $\pm$0.5                   &$17.3^{+0.4}_{-0.3} $ &   \vline &           \\   
\hline
\end{tabular}
}
\caption{UVOT observations of GRB~050802 given in (aperture corrected) count rates and magnitudes. Count rates were aperture corrected to 6$\arcsec$ for V, B and U filters and to 12$\arcsec$ for UVW1, UVM2, UVW2 filters to enable the use of the UVOT zero points to convert the count rates to magnitudes. The zero points are V = 17.83\,$\pm$\,0.09, B = 19.12\,$\pm$\,0.12, U = 18.34\,$\pm$\,0.23, UVW1 = 17.82\,$\pm$\,0.02 ,UVM2 = 17.19\,$\pm$\,0.23, UVW2 = 17.82\,$\pm$\,0.02, White = 19.78\,$\pm$\,0.02.}
\label{tab:uvotpoints}
\end{table*}	

\begin{table*}
\begin{center}
\begin{tabular}{l c } 
\hline
Filter & Ratio\\
\hline
B       &     0.39\\
U       &     0.45\\
UVW1    &     1.71\\
UVM2    &     14.46\\
UVW2    &     12.45\\
White   &     0.13\\
\hline
\end{tabular}
\end{center}
\caption{For 6 UVOT filters, the normalization factor required to convert to V count rate. For B, U, UVW1, UVM2 and UVW2 filters, the normalization factor was calculated by taking the average count rate through the filter in the 400s\,-\,1000s range, and dividing this value by the average count rate in the V filter over the same time interval. The average count rates were then used to create an optical/UV SED, which was used to predict the ratio of White to V count rate.}
\label{tab:uvotnorm}
\end{table*}

\section{Results}

\begin{figure*}
 \includegraphics[angle=-90,scale=0.6]{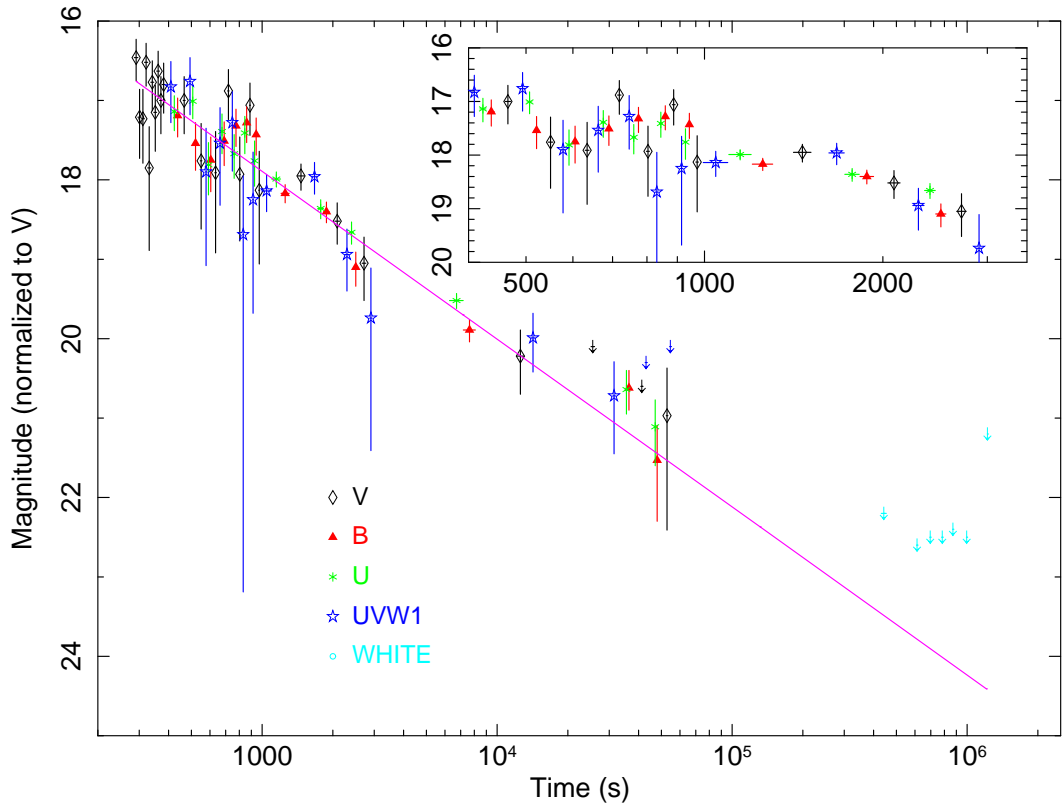}
\caption{Optical Lightcurve of GRB~050802 in 4 of the 6 filters available. The UVM2 and UVW2 filters were excluded from all plots as the data was not constraining. The count rates inside the 10\,s exposures, between 400\,s and 1000\,s were summed and averaged for each filter and the rest of the exposures were normalized against this value. The normalized count rate in each exposure was then divided by the average value for the V filter. Upper limits are given to 3$\sigma$.}
\label{fig:lateopticallc}
\end{figure*}

\subsection{X-ray and Optical Lightcurves}

\begin{figure*}
 \includegraphics[angle=-90,scale=0.6]{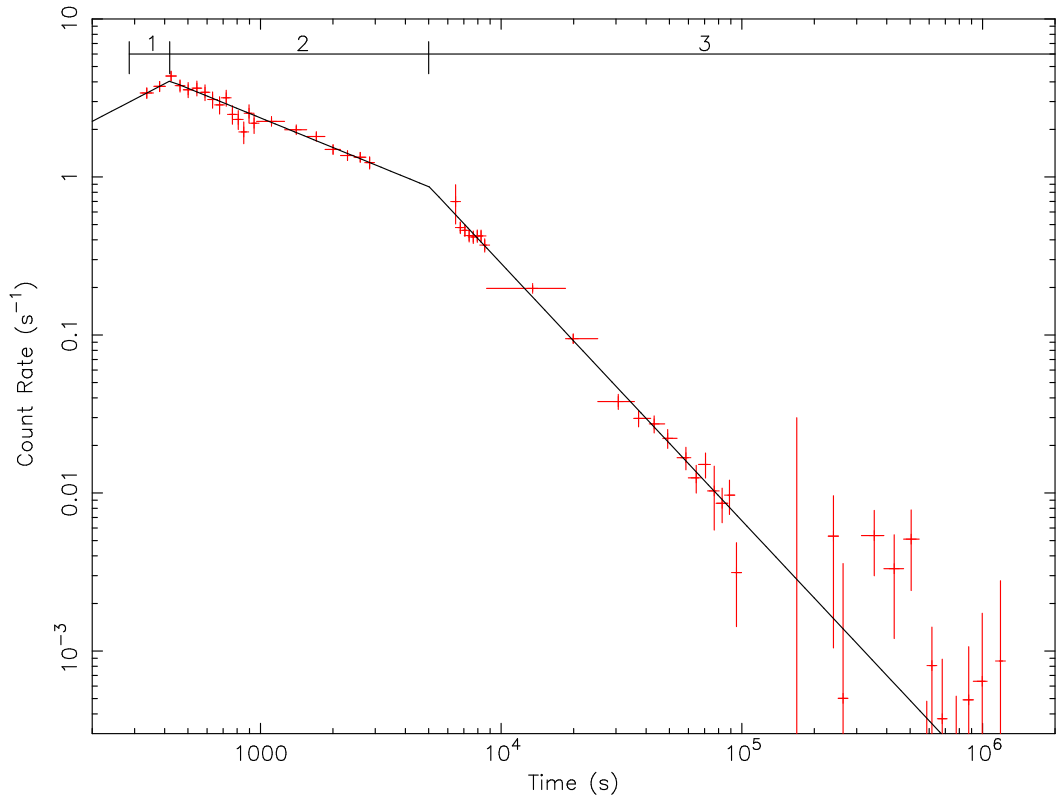}
\caption{X-ray Lightcurve of GRB~050802. A double broken power-law is found to fit the X-ray lightcurve with breaks at 420\,$\pm$\,40\,s and at 5000\,$\pm$\,310\,s.}
\label{fig:xraylc02}
\end{figure*}

The 0.2\,-\,10\,keV X-ray lightcurve of GRB~050802 is shown in Fig.~\ref{fig:xraylc02}. A visual inspection shows a complex behaviour with an initial rise followed by a flat and then a more rapid decay. The X-ray lightcurve was first modelled using a broken power-law. The best fit parameters were $\alpha_1\,=\,0.55\,\pm\,0.03$, $\alpha_2\,=\,1.59\,\pm\,0.03$ and break time 4600\,$\pm$\,260\,s. However the $\chi^2$/D.O.F\,=\,81/57 corresponds to a null hypothesis probability of only 0.02, and the model systematically deviates from the observed lightcurve at the earliest times. Hence, the lightcurve was modelled using a double broken power-law (i.e. a model with 3 power-law segments). This provides a better fit with $\chi^2$/D.O.F\,=\,64/55; according to the F-test the 3-segment power-law fit gives an improvement at the 3$\sigma$ confidence level with respect to the 2-segment model. The values of the best fit parameters are shown in Table \ref{tab:xrayindices}.
	 
In the best fit model, the X-ray lightcurve first rises with a slope $\alpha_1\,=\,-0.80^{+0.71}_{-0.35}$ until 420\,$\pm$\,40\,s. At this point, the lightcurve breaks for the first time and a shallow decay begins with $\alpha_2$\,=\,0.63\,$\pm$\,0.03. This phase ends at $5000\,\pm$\,300\,s when the lightcurve starts to decay steeply with $\alpha_3$\,=\,1.59\,$\pm$\,0.03.

\begin{table*}
\begin{tabular}{|l|c|c|c|c|} 
\hline
Segment & Time at which slope breaks(s)& Energy index $\beta$ &$ N_H (10^{20}$cm$^{-2}$) & Temporal Index $\alpha$\\
\hline
1 (initial rise) &        420 $\pm$ 40    & 0.87 $\pm$ 0.08       & 28 $\pm$ 10  & -0.80$^{+0.71}_{-0.35}$ \\
2 (shallow decay) &        5000 $\pm$ 300 & 0.89 $\pm$ 0.04       & 31 $\pm$ 5    & 0.63$\pm$ 0.03 \\
3 (steep decay)&            -           & 0.88 $\pm$ 0.04       & 28 $\pm$ 5  & 1.59 $\pm$ 0.03  \\
\hline
\end{tabular}
\caption{Spectral and temporal analysis of the X-ray lightcurve of GRB~050802 fitted with 3 segments. The value of $N_H$ refers to the absorption at $z\,=\,1.71$.}
\label{tab:xrayindices}
\end{table*}

\begin{figure*}
 \includegraphics[angle=0,scale=1.0]{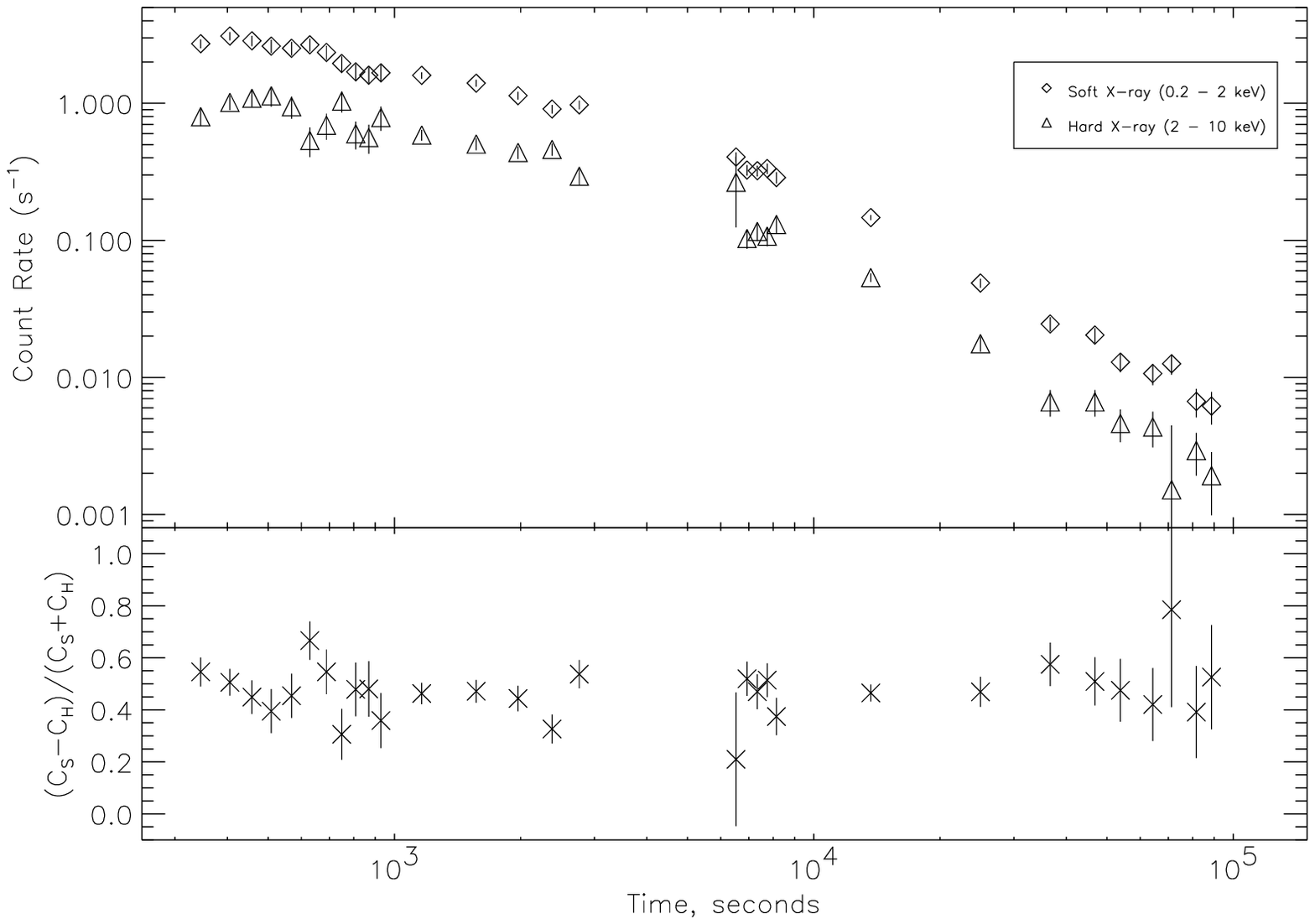}
\caption{Top panel: The soft (0.2\,-\,2keV) and hard (2\,-\,10keV) X-ray lightcurves. Bottom panel: The softness ratio, defined as ($C_S-C_H)/(C_S+C_H$), where $C_H$ is the hard X-ray count rate and $C_S$ is the soft X-ray count rate. There is no evidence for spectral evolution within this time period as can be seen by the constant ratio.}
\label{fig:softness}
\end{figure*}

To look for spectral variations over the course of the decay, we split the lightcurve into soft (0.2\,-\,2\,keV) and hard (2\,-\,10\,keV) X-ray lightcurves. The soft and hard X-ray lightcurves are shown in Fig.~\ref{fig:softness}, where we also show the softness ratio, which we define as ($C_{S}-C_{H})/(C_{S}+C_{H}$) where $C_{S}$\,is the soft X-ray count rate and $C_{H}$\,is the hard X-ray count rate. A constant can be fitted well through the softness ratio time series ($\chi^2$/D.O.F\,=\,30/29) revealing that there is no significant spectral evolution with time. 

The optical lightcurve is shown in Fig.~\ref{fig:lateopticallc} and it is well fitted with a single power-law decay with a temporal index $\alpha_1$\,=\,0.82\,$\pm$\,0.03 ($\chi^2$/D.O.F\,=\,71/63). The lightcurve has no obvious colour evolution within the wavelength range of the UVOT over the duration of the burst afterglow.

\citeauthor{fyn05} (2005a, 2005b) observed an extended source within an arcsecond of the afterglow location, 1.5\,days after the burst trigger. They suggested that it might be the host galaxy and they provided a combined magnitude for the afterglow and extended source of R\,=\,22.5. We have looked for the host galaxy by coadding the UVOT white filter images taken at late times after the afterglow has faded beyond detection. The UVOT white filter observations were taken until $1.2\,\times\,10^6$\,s and are shown in Fig.~\ref{fig:lateopticallc}. The summation of the white exposures provides a deep 3$\sigma$ upper limit of 23.4\,mag, which is equivalent to V\,=\,23.5\,mag for the afterglow spectrum. From these observations, we can confirm that there is no significant contribution from the host galaxy to the lightcurve in the UVOT spectral range for the first 60\,ks, while the afterglow is still detected by the UVOT. 

To determine the earliest time at which the optical lightcurve could have broken, a 2-segment power-law was fitted with the second segment decay rate set to be the same as the third segment in the X-ray lightcurve. We determined the 3$\sigma$ lower limit for the break time by adjusting the time of the break until we obtained  $\Delta \chi^2$\,=\,9 with respect to the single power-law fit. The lower limit to the break time was found to be 19\,ks after the burst trigger, significantly later than the second X-ray break. 

\begin{figure*}  
 \includegraphics[angle=0,scale=1.0]{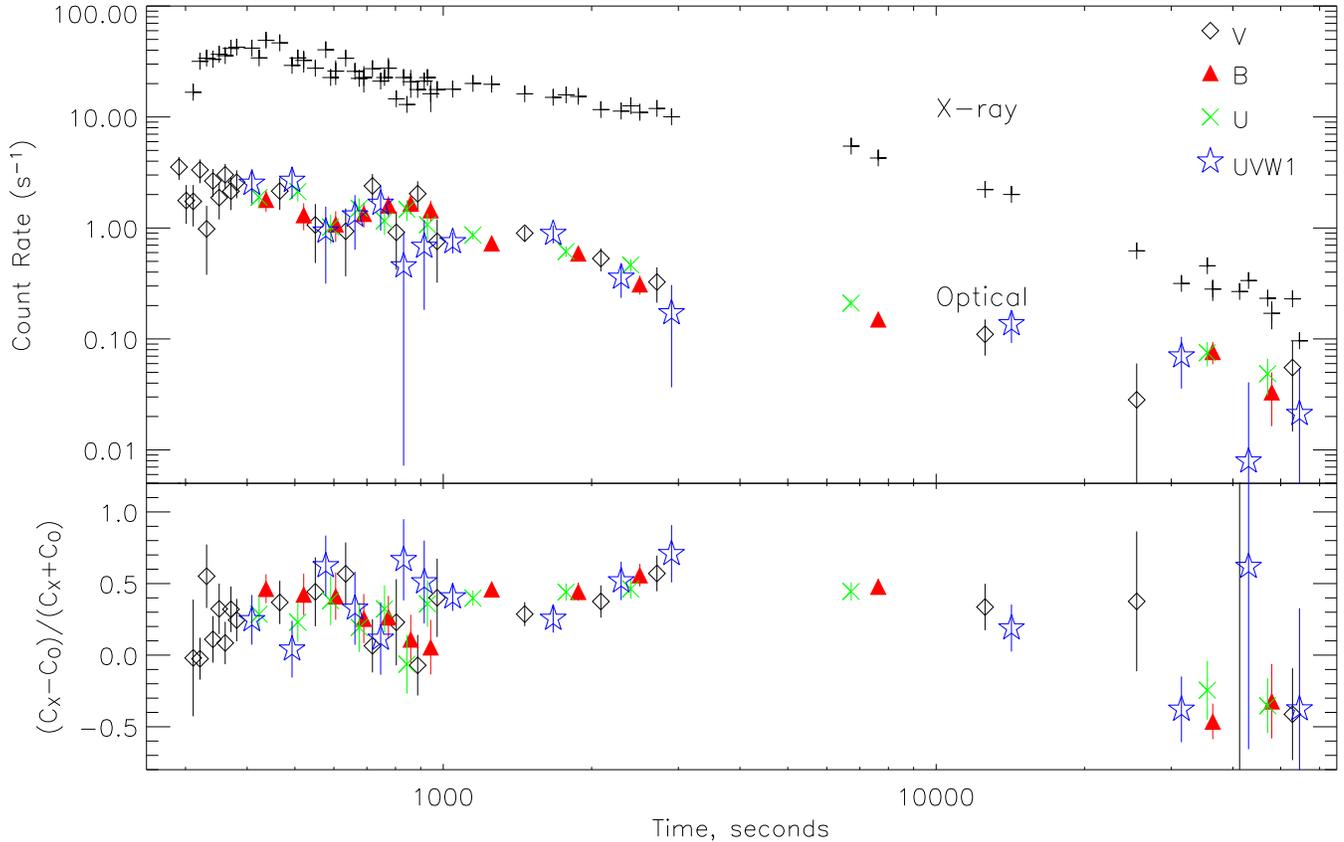}
\caption{Top panel: the X-ray and optical lightcurves. The optical light curve is shown for 4 out of the 6 filters. The count rates in each filter have been normalized to the V filter, as discussed the caption to Fig. \ref{fig:lateopticallc}. The X-ray lightcurve has been binned according to the length of each exposure in the optical lightcurve. Different symbols refer to the filters used for the optical observation in each time bin. Bottom panel: the ratio between the X-ray ($C_X$) and optical ($C_O$) lightcurves.}
\label{fig:opticalxrayratio}
\end{figure*}

In Fig. \ref{fig:opticalxrayratio} we show the X-ray/optical ratio, which we define as ($C_X-C_O)/(C_X+C_O$) where $C_O$\,is the optical/UV count rate normalized to the V filter and $C_X$\,is the X-ray count rate. Initially, we tried fitting a constant across the entire time range. This provided a poor fit with $\chi^2$/D.O.F\,=\,168/62, implying that the changes in the X-ray/optical ratio are highly significant. To investigate the behaviour of the X-ray/optical ratio we fit a function to it before and after the second X-ray break time of 5000\,s. The first 5000\,s was fit by a linear relationship between ($C_X-C_O)/(C_X+C_O$) and log(t) with a gradient of $0.24\,\pm\,0.01$ ($\chi^2/D.O.F\,=\,47/47$). After 5000\,s, the best fit linear relationship has a $\chi^2$/D.O.F\,=\,7/12 and gradient of -0.09\,$\pm$\,0.01. The evolving X-ray/optical ratio implies that the optical/UV to X-ray SED is changing throughout the afterglow.

\subsection{X-ray and Optical Spectra}
The results of the X-ray spectral analysis are shown in Table~\ref{tab:xrayindices}. All spectra were fitted using an absorbed power-law model. The absorption component includes both photoelectric absorption from our Galaxy and from the host galaxy of the GRB. The Galactic column density was fixed at $N_H\,=\,1.78\,\times\,10^{20}\,$cm$^{-2}$ \citep{dic90} and the column density at $z\,=\,1.71$ was allowed to vary. The spectral slopes and column densities measured for the 3 segments of the X-ray afterglow show no evidence of evolution. There is evidence for absorption from the host galaxy of the GRB in each segment; the column density is consistent between the 3 segments of the X-ray lightcurve, with an average value of 2.9$\times\,10^{21}$cm$^{-2}$.
  
\begin{figure*}
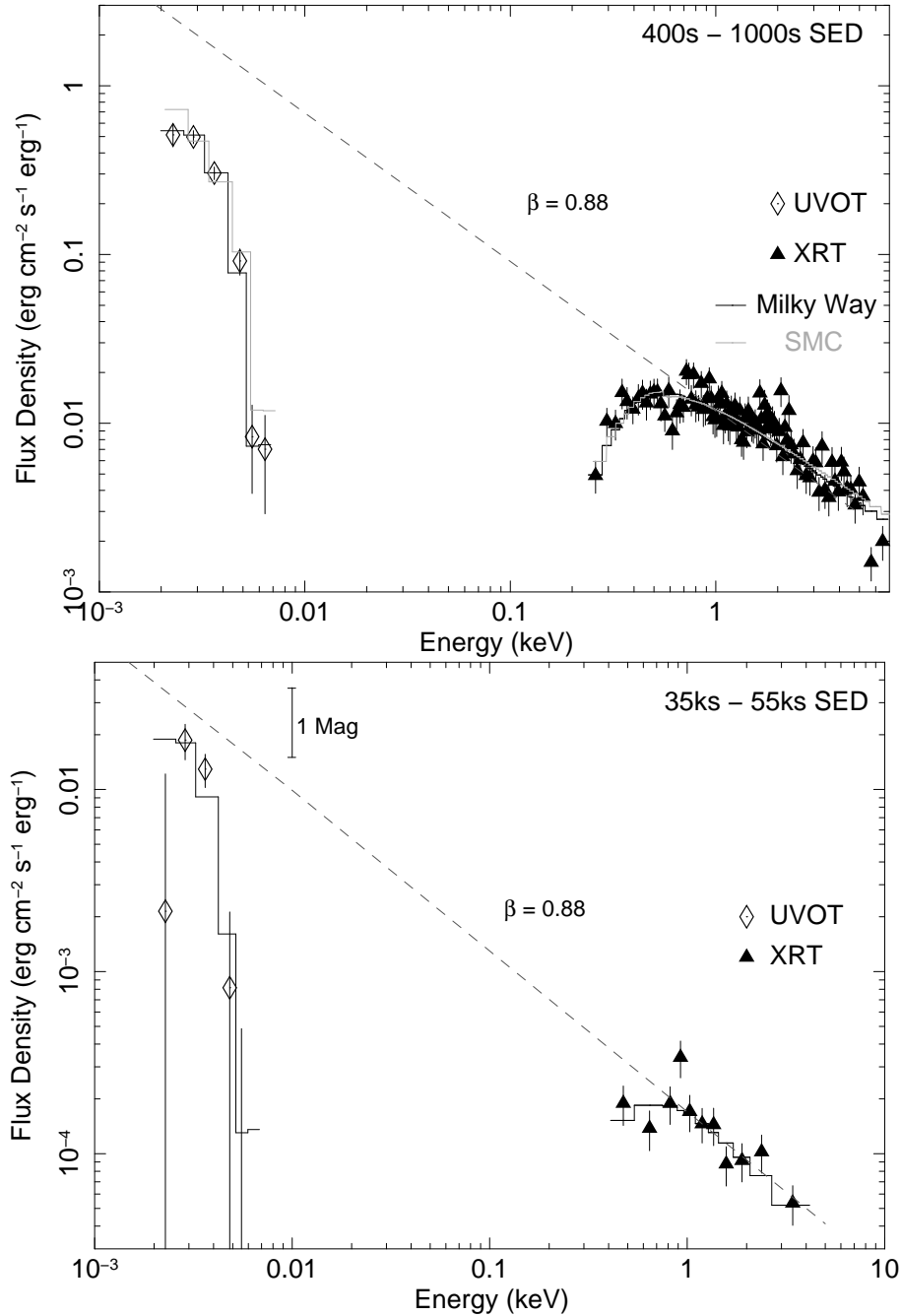

 \includegraphics[angle=-90,scale=0.5]{2ndSED.ps}
\includegraphics[angle=-90,scale=0.5]{3rdsed.ps}
\caption{Optical and X-ray SEDs between 400s and 1000s (top) and 35ks\,-\,55ks (bottom). The top panel shows the fit using SMC dust extinction (grey; offset by a factor of 1.1) and Milky Way extinction (black). The first 6 points in each panel show the 6 filters of the UVOT ranging from V to UVW2. Points after 0.2\,keV show the X-ray spectrum from 0.2\,-\,10\,keV. In the top and bottom panels, the dahsed line represents the weighted mean of the spectral indices found in the 3 segments of the X-ray lightcurve, $\beta$\,=\,0.88.}
\label{fig:SEDearly}
\end{figure*}

Optical/UV to X-ray SEDs in the time intervals 400s\,-\,1000s and 35ks\,-\,55ks were created as described in Section 3.3. For each SED, a power-law fit accounting for Galactic and GRB host-galaxy dust and photoelectric absorption was applied (see Fig.~\ref{fig:SEDearly}). For extinction in our Galaxy, a fixed dust component was used with E(B-V)\,=\,0.03, using the Milky Way extinction curve \citep{pei92}. For extinction in the host galaxy the fit was tried using the SMC extinction curve and then with the Milky Way extinction curve. The fitting was applied to the 400s\,-\,1000s SED as these data have better signal to noise than at late times. The fit using the SMC extinction curve produced $\beta\,=\,0.79\,\pm\,0.02$ and $\chi^2$/D.O.F\,=\,134/104. The fit using the Milky Way extinction curve returned $\beta\,=\,0.86\,\pm\,0.02$ and $\chi^2$/D.O.F\,=\,120/104. We also tested a model in which a cooling break resides in between the optical and X-ray bands. In this case, a broken power-law model was tested in which the spectral indices have a fixed difference of $\Delta\beta\,=\,0.5$. For consistency, the model was tried using the SMC and MW extinction curves. The broken power law using the SMC extinction curve returned $E_{break}\,=\,0.010^{+0.009}_{-0.008}$keV and $\beta_2\,=\,0.89\,\pm\,0.01$ with $\chi^2/D.O.F\,=\,125/103$. The fit using the Milky Way extinction curve gave $E_{break}\,=\,0.004^{+0.005}_{-0.003}$ and $\beta_2\,=\,0.89\,\pm\,0.04$ with $\chi^2/D.O.F\,=\,119/103$. Overall, the fits with the Milky Way extinction curve provide the best $\chi^2/D.O.F$ and for this extinction curve there is no significant improvement to the fit by replacing a power-law with a broken power-law. The model parameters from the different model fits and the implied total (Galactic and GRB host galaxy) extinction in the UVOT bands are given in Table \ref{tab:extinction}.

Unusually, we find that the Milky Way extinction curve best fits the SED of GRB~050802. In comparison, no other bursts in the samples of \cite{sch07} or \cite{sta06}, which consist of 7 {\it Swift} GRBs and 10 {\it BeppoSAX} GRBs respectively, are fitted best with a Milky Way extinction curve. Since the extinction curve is unusual, we determined the $N_H/A_V$ to see how this compares to other GRBs. For GRB~050802, the $N_H/A_V$ ratio was found to be $4.5\,(\pm 2.3)\,\times\,10^{21}$. The mean GRB $N_H/A_V$ ratio for a MW extinction law in the sample of \cite{sch07} is 4.7$^{+1.4}_{-1.3}\times\,10^{21}$, so GRB~050802 is consistent with the mean $N_H/A_V$ in \cite{sch07} to within 1$\sigma$. This implies that the ratio of dust and gas surrounding GRB~050802, is fairly typical for GRBs. 

Because we have shown that the absorption does not change significantly with time, we do not expect the extinction to change either. Therefore, in fitting the 35ks\,-\,55ks SED we froze the extinction and absorption at the best fit values found for the 400s\,-\,1000s SED. This resulted in a fit with $\beta\,=\,0.99\,\pm 0.02$ ($\chi^2$/D.O.F\,=\,27/15), which implies a null hypothesis of only 3$\%$. Furthermore, the value of $\beta$ obtained in this fit is inconsistent at 99 per cent confidence with $\beta_3\,=\,0.88\pm$\,0.04, the spectral index of the third segment of the X-ray lightcurve. If we repeat the fit to the 35ks\,-\,55ks SED with a fixed spectral index of $\beta$\,=\,0.88, we obtain a $\chi^2/D.O.F\,=\,35/16$, which implies the model is rejected at 99 per cent confidence. 

\begin{table*}
\begin{tabular}{lllll@{\hspace{10mm}}l}
\hline
Model Parameters &\multicolumn{4}{l}{-------------------------------- Models for 400s\,-\,1000s SED --------------------------------}&35ks\,-\,55ks SED\\
\& Host Extinction &Power-law&Power-law&Broken power-law&Broken power-law&Power-law\\
&MW extinction& SMC extinction& MW extinction&SMC extinction& MW extinction \\
\hline
$\beta$&       0.86\,$\pm$\,0.02&0.79\,$\pm$\,0.02  & 0.89\,$\pm$\,0.04&    $0.89\,\pm\,0.01$   &0.99\,$\pm$\,0.02  \\
$ Energy_{break}$ (keV)& - & - &0.004$^{+0.005}_{-0.003} $& 0.010$^{+0.009}_{-0.008}$ & -  \\
$ E(B-V)_{host}$    &   0.18\,$\pm$\,0.02& 0.09\,$\pm$\,0.01  & 0.18\,$\pm$\,0.02&0.10\,$\pm$\,0.02 & 0.18*\\
$ N_{H_{host}} $($\times10^{20}$cm$^{-2})$ & 26\,$\pm$\,4 &20\,$\pm$\,4 &29 $\pm$\,6& 30\,$\pm$\,4 & 26*\\
$ A_V$ (Mag)&        1.6$\pm$0.2 & 0.8$\pm$0.1 & 1.6$\pm$0.2 & 0.8$\pm$0.1  & 1.6* \\
$ A_B$ (Mag)&        1.4$\pm$0.2 & 1.2$\pm$0.1 & 1.5$\pm$0.2 & 1.1$\pm$0.1  & 1.4* \\
$ A_U$ (Mag)&        1.7$\pm$0.2 & 1.4$\pm$0.2 & 1.8$\pm$0.2 & 1.4$\pm$0.2  & 1.7* \\
$ A_{UVW1}$ (Mag)&   2.5$\pm$0.3 & 1.8$\pm$0.2 & 2.6$\pm$0.3 & 1.8$\pm$0.3  & 2.5* \\
$ A_{UVM2}$ (Mag)&   2.8$\pm$0.3 & 1.9$\pm$0.2 & 2.9$\pm$0.3 & 1.9$\pm$0.3  & 2.8* \\
$ A_{UVW2}$ (Mag)&   2.8$\pm$0.3 & 1.9$\pm$0.2 & 2.9$\pm$0.3 & 1.9$\pm$0.3  & 2.8* \\ 
$\chi^2$/D.O.F&120/104 &134/104&119/103 &125/103 &27/15\\
\hline
\end{tabular}
\caption{The model parameters were determined from the fitting of a power-law and separately a broken power-law to the 400s\,-\,1000s SED; the fits were repeated for the MW and SMC extinction curves. Also shown, is the power-law with MW extinction fit to the 35ks\,-\,55ks SED. Parameters marked with * are fixed at the best fit values found for the power-law fit to the 400s\,-1000\,s using MW extinction. All models use a fixed value of the Galactic extinction $E(B-V)$\,=\,0.03 and Galactic absorption $N_H\,=\,1.78\,\times\,10^{20}\,$cm$^{-2}$. The observed host extinction in each filter is provided for each fit; note that these were derived from the fit parameter $E(B-V)_{host}$ and were not fit parameters in their own right.}
\label{tab:extinction}
\end{table*}

\section{Discussion}

In summary, we have determined that the X-ray lightcurve can be divided into 3 segments with two breaks at 420\,s and 5000\,s. Throughout the afterglow, there is no evidence for X-ray spectral evolution. We also find that the optical lightcurve decays as a single power-law, which has a slightly steeper decay than the second segment of the X-ray lightcurve. We have determined the earliest possible time that the optical lightcurve could have changed its decay rate to match that of the third X-ray segment to be 19\,ks, which is 14\,ks after the last X-ray break. 

In the following subsections, we will look in detail at the X-ray afterglow to examine the origin of the X-ray breaks and to determine the best fitting closure relations between the temporal and spectral slopes. Then we shall discuss the question of why the X-ray lightcurve breaks at 5000\,s while the optical lightcurve continues to decay as a power-law. Several mechanisms will then be investigated to find a compatible model for the production of the X-ray and optical afterglow.

\subsection{The X-ray Afterglow}
We begin by looking at the first break to see what we may learn. The break occurs at $420\,\pm\,40$\,s, which is preceded by a rise from 286\,s with a slope of -0.80$^{+0.71}_{-0.35}$. The break is then followed by a gradual decay with a slope of $\alpha\,=\,0.63\,\pm\,0.03$. The rise and the successive slow decay can be regarded as a broad peak in the lightcurve and it may be attributed to `flaring' activity \citep{bur05b,nou06} or could be the early phase of the jet interacting with the external medium, giving rise to the forward shock \citep{kum00}. However, it is not possible to discriminate between the two mechanisms. Flares often have different spectra to the afterglow, but if this is a flare it may be so small we would not expect to see any spectral variability. On the other hand, we can not tell if this is the early rise of the afterglow as we have too little data prior to the peak. An additional contribution to the second segment from a large flare may also be excluded as a power-law fit to the X-ray lightcurve between 1000s and 3000s results in a slope of $\alpha\,=\,0.64\,\pm\,0.05$ with $\chi^2$/D.O.F\,=\,21/38. 

Typically, X-ray afterglows are analysed by applying the closure relations given by \cite{zhaf05} to the X-ray spectral and temporal indices. These relations are an aid in determining the location of the observed X-ray band relative to the synchrotron frequencies ($\nu_a,\nu_m,\nu_c$) and the environment in which the the burst occurs. 

We skip the first segment of the X-ray lightcurve as the closure relations are not applicable to a lightcurve that is rising. We begin by applying the simplest non-injected relations to the second segment, in which $\alpha_2\,=\,0.63\,\pm\,0.03$ and $\beta_2\,=\,0.89\,\pm\,0.04$, the only closure relation that agrees without the requirement of energy injection at 3$\sigma$ confidence level, is \begin{equation}\alpha\,=\,\frac{3\beta-1}{2}\label{equ:2}\end{equation} This relation applies to a wind or ISM medium with $\nu_X\,>\,\nu_c$, but, this relation is not appropriate for GRB~050802 as the value of the energy index $p$, determined through $\beta$\,=\,$p$/2 is $p$\,=\,1.78$\,\pm$\,0.08. Values of $p$\,$<$\,2 require a different set of closure relations provided by \cite{zha04a}, of which none satisfies the second decay of GRB~050802. More complex closure relations, which include a third parameter to account for energy injection, are provided by \cite{zhaf05}. In these models the luminosity is assumed to evolve as L(t)\,=\,$L_0(t$/$t_b)^{-q}$ where $q$ is the luminosity index and $t_b$ is the time for the formation of a self-similar solution, which is approximately the deceleration time. The luminosity index $q$ must be \,$<$\,1 for the energy injection to affect the afterglow dynamics. We found that there are two injected closure relations that are consistent with $\alpha_2$ and $\beta_2$. The first relation is \begin{equation}\alpha\,=\,\frac{q}{2}+\frac{(2+q)\beta}{2}\label{equ:3}\end{equation} This closure relation is for a wind medium with slow cooling electrons with either $\nu_X<\nu_m$ or $\nu_m<\nu_X<\nu_c$ and where $q$\,=\,$-0.27\,\pm\,0.04$. The second possible energy injected relation is \begin{equation}\alpha\,=\,(q-1)+\frac{(2+q)\beta}{2}\label{equ:4}\end{equation} This relation is for a uniform medium with slow cooling electrons with either $\nu_m\,<\,\nu_X$ or $\nu_m\,<\,\nu_X\,<\,\nu_c$ and where $q$\,=\,$0.51\,\pm\,0.04$. For the closure relations given in Equations \ref {equ:3} and \ref{equ:4}, $\beta\,=\,(p-1)/2$ and so $p\,=\,2.79\pm\,0.08$.

When the closure relations are applied with the indices of the third segment, where $\alpha_3\,=\,1.59\,\pm\,0.03$ and $\beta_3\,=\,0.88\,\pm\,0.04$, we find that the most consistent closure relation without energy injection is \begin {equation}\alpha\,=\,(3\beta+1)/2 \label{equ:1}\end{equation}  which is consistent at 3.5$\sigma$ confidence. All other non-injected closure relations are ruled out at a minimum of 4$\sigma$. Equation \ref{equ:1} is for electrons that are slow cooling within the range $\nu_m$\,$<$\,$\nu_X$\,$<$\,$\nu_c$ in a wind medium. Again a value for the energy index may be determined though $\beta\,=\,(p-1)/2$ with $p$\,=\,2.76\,$\pm$\,0.08. Generally, the energy index is expected to remain constant throughout the X-ray afterglow, unless it is changed by energy injection \citep{zhaf05}.

Therefore, the application of the closure relations to the X-ray temporal and spectral indices suggests that the second segment is energy injected and the third segment is not, and therefore that the break between the second and third segments at $\simeq$\,5000\,s is due to the discontinuation of energy injection. At this point it is worth pointing out that the second and third segments of GRB~050802 are consistent with the second and third segments of the canonical lightcurve presented by \cite{nou06}. Furthermore, from the application of the closure relations above we would come to the same physical explanation as \cite{nou06} and \cite{zhaf05} for this temporal behaviour: that the break between the two segments corresponds to the end of energy injection.
Having looked at the X-ray afterglow in isolation, we will now investigate whether this physical picture is compatible with the observed optical emission. The standard afterglow is based on the assumption of a single relativistic outflow \citep{mes97} and we will begin by considering models of this type before considering multi-component outflows.

\subsection{Can the X-ray and optical afterglow be explained by a single jet?}
We start by considering an afterglow in which the emission in both the X-ray and optical bands is produced predominantly by the forward shock. If we compare the decay of the optical lightcurve with the segments of the X-ray lightcurve we find that the optical decay is similar, although not identical to, the slow `energy injected' decay of the second segment. This would suggest that the optical lightcurve is also characteristic of an energy-injected afterglow. Since the X-ray lightcurve breaks at 5000\,s, suggesting the end of energy injection, and the optical lightcurve continues with a shallow decay indicative of energy injection, we first consider the possibility that the electrons responsible for producing the optical emission continue to receive energy while the X-ray emitting electrons do not. 

There are two possible mechanisms for continued energy injection. These are the arrival of additional shells of material at the shock region or the release of energy stored in the form of Poynting flux. In the first case, the injected energy changes the fireball dynamics, but can only change the balance of X-ray to optical emission and produce a break in one of the lightcurves if $\nu_m$ or $\nu_c$ pass through one of these bands. However, we can rule out the passage of one of these breaks through the X-ray or optical afterglow because there is no break in the optical lightcurve and there is no change in the X-ray spectrum. This therefore rules out continued energy in the form of blast waves and so we consider continued energy injection in the form of Poynting flux. Energy is supplied in the form of Poynting flux when there is a rotating compact object that has a magnetic field, $B$. The flux is converted to kinetic energy when kink instabilities cause the magnetic field lines to reconnect in the outflow \citep{dre02,gia06}. The energy density of Poynting flux is $\propto B^2$, and  within the reconnection region the energy available per particle for particle acceleration is $\propto B$ \citep{gia06b}. To decrease the photon energy so that further emission is below the X-ray band, requires a decrease of the average particle energy. The XRT is sensitive over more than a decade in photon energy, therefore a decrease in the average photon energy from above the XRT energy band to below the XRT energy band requires a factor $>$10 change in $B$ and therefore a factor of $>100$ change in the energy density of Poynting flux \citep{gia06b}. This implies a large change in the overall rate of energy injection, which is inconsistent with the continued power-law decay in the optical.

A reverse shock is an alternative means for producing optical emission without X-ray emission and so we now examine the possibility that this could sustain the optical emission of GRB~050802 after the X-ray lightcurve has broken. The reverse shock will travel back through the ejected matter and will emit as long as it is passing through ejecta. The reverse shock emission will cease when it passes over the last ejected shell \citep{zha02}. The shock crossing time is expected to be approximately the duration of the GRB, but if energy continues to be injected into the afterglow by the arrival of further shells the reverse shock may continue for as long as the energy injection phase. If the end of energy injection occurs at 5000\,s, as suggested by the X-ray lightcurve, then we should expect that the reverse shock must cease at approximately 5000\,s. Therefore, this is not the required mechanism as the optical lightcurve has no break until at least 19\,ks.

\cite{pan06} have proposed that it is possible to sustain a power-law decay in the optical, while the X-ray lightcurve shows a break due to the end of energy injection, if the microphysical parameters are changing with time. The microphysical parameters that must change to keep the optical temporal decay as a power-law are: the fraction of post-shock energy in the magnetic fields $\epsilon_B$, the fraction of post-shock energy that is given to electrons $\epsilon_i$ and the blast-wave kinetic energy E \citep{nou06,pan06}. These parameters are expected to evolve with Lorentz factor $\Gamma$, if they evolve at all. In the scenario proposed by \cite{pan06}, the cooling frequency $\nu_c$ lies between the optical and X-ray bands. In this case, there should be a slope change of $\Delta\beta\,=\,0.5$ between the X-ray and optical bands and $\nu_c$ moves to lower frequencies with time \citep{pan06}. For GRB~050802, through the fitting of a broken power-law with a fixed $\Delta\beta\,=\,0.5$ to the 400s\,-\,1000s SED (see Section 4.2), the maximum energy, at which a break could be present, was determined to be \,0.02\,keV. 

We examined the optical to X-ray SEDs to see if the X-ray and optical lightcurves can be explained by $\nu_c$ moving to lower frequencies with time as expected in the model of \cite{pan06}. In this model the largest change in the ratio $(C_X-C_O)/(C_X+C_O)$, where $C_X$ is the X-ray count rate and $C_O$ is the V band count rate corresponds to the motion of $\nu_c$ from its highest allowed value to below the optical pass band, at which point the optical emission lies on the same power-law spectral segment as the X-ray emission. In the broken power-law fit to the 400s\,-\,1000s SED, the highest allowed value of $\nu_c$ at 3$\sigma$ confidence is 0.06\,keV. Therefore we refited the  400s\,-\,1000s SED with $\nu_c$ fixed at 0.06\,keV, then reduced the value of $\nu_c$ to below the UVOT spectral range and so determined that the smallest $(C_X-C_O)/(C_X+C_O)$ value allowed by this model is $-0.10$. The observed range of $(C_X-C_O)/(C_X+C_O)$ extends to much lower values than this (Fig. \ref{fig:opticalxrayratio}). We find that the weighted mean of the data points more than $35$\,ks after the BAT trigger is $(C_X-C_O)/(C_X+C_O)\,=\,-0.38\,\pm\,0.08$, which is $>3\sigma$ below the lowest allowed value of $-0.10$, thus allowing the model of \cite{pan06} to be ruled out.

 We now look in to the optical to X-ray SED further and ask: to what extent can single component models be ruled out completely? As discussed in Section 4.2, a power-law fit to the late time (35ks\,-\,55ks) SED provides a poor fit and the best fit value of $\beta$ is softer than the X-ray spectral index for segment 3, which includes the 35ks\,-\,55ks time interval. In this respect, we note that there is no evidence for any evolution in the X-ray spectral index at any point in the afterglow and the weighted mean of the indices found in the 3 segments of the X-ray lightcurve gives $\beta$\,=\,0.88\,$\pm$\,0.03. To see why the fit to the late time SED is poor we show the X-ray spectrum extrapolated to lower energies using this spectral slope ($\beta$\,=\,0.88) as a dashed line in both panels of Fig \ref{fig:SEDearly}. The dashed line lies above the optical and UV data points as expected for an afterglow with significant extinction. In Section 4.2, we determined that at $3\sigma$ confidence, the host galaxy extinction is greater than 0.8 and 1.1 in the B and U bands respectively, no matter which models we choose for the extinction and continuum. In addition, the Galactic extinction is 0.1 for both the B and the U bands. After correction for extinction, the B and U band fluxes in the late time SED lie significantly above the dashed line. As described in Section 4.2, this power-law can be ruled out with $>$99 per cent confidence, implying at 99 per cent confidence that the region responsible for the X-ray emission can not produce all of the optical emission at late times, no matter where $\nu_c$ and $\nu_m$ lie. Since single component outflows are unable to represent the late time SED satisfactorily, we now consider multi-component outflows to see which of these models are able to reproduce the observed lightcurves and SEDs.

\subsection{Multi-component Outflows}
A multi-component outflow is one consisting of two or more components that have different bulk Lorentz factors. The simplest model is one of two components: a narrow jet and a second wider, but slower jet that surrounds the narrow component. This geometry may be generated by the narrow jet giving rise to a wider and slower component \citep{kum03}, where the bulk $\Gamma$ decreases over time and produces a jet with angular structure. Alternatively, the two components may be formed at the same time for example when neutrons and protons decouple in a neutron-rich, hydroydynamically accelerated jet from a neutron star or from the neutron-rich accretion disk of a collapsed massive star \citep{vla03}. \cite{pen05} show that within such a neutron-rich jet, the wide component has greater energy than the narrow component so that the wider component is able to dominate at late times. In both cases, the narrow component will produce the X-ray emission, whereas the optical emission will be produced by the wider component as it travels at lower Lorentz factors than the narrow component. This picture can be used to explain bursts such as GRB~021004 and GRB~030329 \citep{pen05}. The wider jet is expected to reach the undisturbed medium at a slightly later time than the narrow jet. Thus, the wide component may be observed as a later start for the optical lightcurve when compared to the X-ray lightcurve, or on an established optical lightcurve, the arrival of the wide jet to the external medium may be shown as a late rise/bump or a change in the decay slope. 
 
There are three options within a two-component jet we will investigate to see if one could have produced the observed lightcurves of GRB~050802. All of these options involve energy injection, which is introduced through shells of material. The options are: energy injection with the ejected shells wide enough to refresh both components of the jet with a Lorentz factor that decreases over time (so that after 5000\,s the shells no longer reach the shock in the narrow component), the cessation of energy injection at 5000\,s in one component only and, finally, the continuation of energy injection for the duration of the afterglow, but with the narrow component producing a jet break at 5000\,s. 

The first option is that energy injection is distributed uniformly throughout both jets and continues for a sustained period of time. Initially, the shells have a wide enough angular distribution, and travel at a $\Gamma$ large enough, to reach the forward shock produced by both components. Both the X-ray and the optical afterglows will appear energy injected. As time goes by, the Lorentz factor of the freshly ejected shells decreases. Eventually, the ejected shells will not travel at a Lorentz factor large enough to reach the shocked region in the narrow component, but will still be able to reach the wide component. This will cause a break to be seen only in the X-ray afterglow. However, the freshly ejected shells will still travel at $\Gamma$ large enough to reach the wide component and therefore the optical lightcurve will continue to appear energy injected. This much appears to be consistent with the data because, as discussed in Section 5.2, the optical lightcurve appears to be energy injected for its entire observed duration and does not break with the X-ray lightcurve. However, as the wide component is slower than the narrow component, the wide component will absorb shells (i.e receive energy) at a faster rate than the narrow component. Therefore, the optical emission is expected to decay less rapidly than the X-ray emission. For GRB~050802, the opposite is actually observed: in the optical the decay index $\alpha_O\,=\,0.82\,\pm\,0.03$, while the X-ray second segment has a decay index of $\alpha_2\,=\,0.63\,\pm\,0.03$, thus this model is excluded. 

 The second option calls for the X-ray break at 5000\,s to have been caused by the cessation of energy injection in the narrow component, while injection continues in the wide component. However, this implies that the energy is injected as a hollow jet at late times and we consider this to be an unattractive solution on physical grounds, and lacking in testable predictions.

The third option is that energy is continuously injected into both components and the X-ray break at 5000\,s results from a jet break of the narrow component \citep{pan04}. The spectral indices of the X-ray segments support this model as there is no spectral change over the break, as is expected for a jet break. A jet break may be produced within a laterally expanding or a non-expanding jet. The closure relations for these cases are provided by \cite{pan06b}. In order for this scenario to be viable we have to find a post jet-break closure relation that has consistent energy injection, external medium, synchrotron frequencies and electron spectral index to that of the pre jet-break closure relation. The closure relation that satisfies this uses Eq 35 of \cite{pan06}, and is:
\begin {equation}\alpha\,=(1+2\beta)-\frac{2}{3}(1-q)(\beta+2) \label{equ:6}\end{equation}
This relation is independent of the GRB host medium and is for a laterally expanding jet with $\nu_m\,<\,\nu\,<\,\nu_c$. The relation provides a value of $q$\,=\,$0.38\,\pm\,0.04$, which is consistent at the 99 per cent confidence level with the value of $q$ obtained through the closure relation in Equation \ref{equ:4}. The consistency with Equation \ref{equ:4} implies that this burst originated in a uniform ISM and not a wind medium as suggested in Section 5.1.

Furthermore, we can use the second X-ray break time to infer the opening angle of the narrow beam and to put a lower limit on the angle of the wide component. We calculated the angles using the following expression from \cite{fra01}:
\begin {equation} \theta_{{\rm jet}}=0.057\  t_{jet}^{3/8}\left(\frac{1+z}{2}\right)^{-3/8}\left(\frac{0.2E_{iso}}{\eta }\right)^{-1/8}\left(\frac{n}{0.1}\right)^{1/8} \end{equation}
 where $E_{iso}$ is the isotropic energy of the burst, $n$ is the density of the medium and $\eta$ is the efficiency of converting energy in the ejecta into $\gamma$-rays. We use values of $\rm E_{iso}$\,=\,0.2$\,\times\,10^{53}$ergs (in the range 15\,-\,350 keV) and GRB efficiency $\eta$\,=\,0.2. Approximating the circumburst medium as a spherical, uniform gas cloud of radius $R$, the particle density $n$, may be estimated simply as $n=N_H/R$. The cloud radius for gas and dust to survive the GRB, may be estimated using Fig. 8 of \cite{per02}. This shows that the obscuring medium must extend to $R\sim 10^{20}$cm, implying a number density of $n\,\sim$\,30 cm$^{-3}$. Thus, the value for the narrow component is estimated to be $\theta_{jet,narrow}\,\sim\,1^\circ$ and for the wide component the opening angle is $\theta_{\rm jet,wide}\,\geq\,8^\circ$. We recognise that these angles are only approximate because they depend weakly on the bolometric energy, the efficiency of the burst and on the particle density of the external medium. 

This model requires that energy be continually injected into both components of the jet for the observed duration of the burst. The time scale over which this energy injection takes place is more than an order of magnitude larger than $10^3-10^4$\,s, which is the duration of energy injection required to explain the canonical X-ray afterglow \citep{nou06,zhaf05,cap06,tas06,kri06}. This situation may seem unusual, but GRB~050802 is not the first burst that has required discrete or continuous energy injection for an extended period. A few examples of bursts that have required long duration energy injection are GRB~021004, XRF~050406 and GRB~060729 \citep{bjo04,roman06,gru06}. The continuation of energy injection well beyond the jet break time will produce shallower post jet break slopes than the non-injected case and these may be mistaken for the discontinuation of energy injection. This would lead to the surprisingly small number of reported jet breaks in the {\it Swift} era \citep{wil06}. However, the mechanism that could be responsible for long duration ($>10^4$\,s) energy injection is uncertain. Currently, the most favoured scenario is one in which the duration of the central engine is short, but where the ejecta are emitted with shells of varying velocities \citep{rees98}. The shells will order themselves before reaching the external medium and will form a continuous flow of shells. In this case, small time scale fluctuations need not be observed in the power-law lightcurves and the central engine is not required to be particularly long lived. 

This model would allow the break to occur only in the X-ray lightcurve because the narrow and wide components travel at different $\Gamma$, and the narrow component is beamed into a smaller angle. Therefore, the two component outflow with continued energy injection and a jet break in the narrow component is our favoured hypothesis for producing the observed lightcurve of GRB~050802.

The result from this paper has wider implications for our understanding of GRBs. Segments two and three of the X-ray lightcurve of GRB~050802 have the same characteristics as the segments two and three of the canonical X-ray afterglow lightcurve \citep{nou06}. If we had looked at the X-ray data alone, then our explanation for the second and third X-ray segments of GRB~050802 would have been equivalent to the explanation proposed by \cite{nou06} and \cite{zhaf05} for the canonical lightcurve: an energy injected decay followed by a `normal' decay. However, when the optical observations are taken into account, this explanation is no longer attractive. Instead, a more plausible explanation is a two component outflow with the break in the X-ray lightcurve likely to be due to a jet break. Our results suggest that without the optical lightcurve the interpretation of GRB afterglows may not be correct.

\section{Conclusion}
This paper has investigated GRB~050802, an unusual burst because the optical decays as a power-law, while the X-ray breaks twice. We have analysed optical and X-ray data gathered from the XRT and UVOT instruments onboard {\it Swift}. Observations continued with both the XRT and the UVOT up until 1.2$\times10^{6}$\,s.

The analysis of the afterglow began by looking at the origin of the three power-law segments in the X-ray lightcurve. The X-ray lightcurve commences with a rise until 420\,s. The second segment decays slowly with $\alpha_2$\,=\,0.63\,$\pm$\,0.03 until 5000\,s. The third and last segment decays faster with a slope of $\alpha_3$\,=\,1.59\,$\pm$\,0.03. The optical lightcurve decays as a single power-law with $\alpha_{O }$\,=\,0.82$\pm$0.03. 

 Through modelling of the 400s\,-\,1000s X-ray to optical SED, it was determined that the best fit is a power-law with Milky Way extinction. However, when extinction in the host galaxy is accounted for, the optical points lie above the X-ray power-law at late times, indicating that the optical afterglow could not be produced by the same component as the X-ray emission. This result rules out single component afterglows with 99 per cent confidence. 

The next logical step was to examine a multi-component outflow. Several variations of a two component jet, the simplest multi-component outflow, were investigated. Of all the possibilities, we find the most physically self-consistent model is one in which both a narrow and wide component are energy injected for the duration of the observed afterglow and the X-ray break at 5000\,s is due to a jet break in the narrow component. 

This paper has wide implications for the GRB community. We have found that the explanation obtained by examining only the X-ray data of GRB~050802 differs remarkably from the answer obtained by examining the optical and X-ray emission together. Our results suggest that without the optical lightcurve the correct interpretation of GRB afterglows may not be possible.
    
\section{Acknowledgments}
SRO acknowledges the support of a PPARC Studentship. SZ thanks PPARC for its support through a PPARC Advanced
Fellowship. We also thank the referee for useful comments and suggestions.

\label{lastpage}
\end{document}